\documentclass[aps,prl,twocolumn,superscriptaddress,longbibliography]{revtex4-2}
\usepackage{amsmath}
\usepackage{amsfonts}
\usepackage{amssymb}
\usepackage{lmodern,dsfont}
\usepackage{graphicx}
\usepackage[usenames,dvipsnames]{xcolor}
\usepackage{bm}
\usepackage[english=american]{csquotes}
\usepackage{xr}

\newcommand{\Av}[1]{\left\langle #1 \right\rangle}
\newcommand{\av}[1]{\langle #1 \rangle}
\newcommand{\n}{\nonumber}
\newcommand{\nn}{\nonumber \\}

\newcommand{\grad}{\bm{\nabla}}

\renewcommand{\eqref}[1]{Eq.~(\ref{#1})}

\begin{document}
\author{Andreas Dechant}
\affiliation{Department of Physics \#1, Graduate School of Science, Kyoto University, Kyoto 606-8502, Japan}
\author{Shin-ichi Sasa}
\affiliation{Department of Physics \#1, Graduate School of Science, Kyoto University, Kyoto 606-8502, Japan}
\author{Sosuke Ito}
\affiliation{Department of Physics, The University of Tokyo, Tokyo 113-0033, Japan}
\affiliation{Universal Biology Institute, The University of Tokyo, Tokyo 113-0033, Japan}
\affiliation{JST, PRESTO, Saitama 332-0012, Japan}

\title{Geometric decomposition of entropy production in out-of-equilibrium systems}
\date{\today}

\begin{abstract}
Two qualitatively different ways of driving a physical system out of equilibrium, time-dependent and non-conservative forcing, are reflected by the decomposition of the system's entropy production into excess and housekeeping parts.
We show that the difference between these two types of driving gives rise to a geometric formulation in terms of two orthogonal contributions to the currents in the system.
This geometric picture in a natural way leads to variational expressions for both the excess and housekeeping entropy, which allow calculating both contributions independently from the trajectory data of the system.
We demonstrate this by calculating the excess and housekeeping entropy of a particle in a time-dependent, tilted periodic potential.
\end{abstract}

\maketitle

A particle system in equilibrium with its environment may be driven out of equilibrium in two qualitatively different ways:
Either, we vary the parameters of the system according to a time-dependent protocol.
In this case, if we imagine suspending the protocol at any given instant, the system will relax back to the equilibrium state corresponding to the instantaneous values of the parameters.
Alternatively, we can also apply a time-independent, nonconservative force to the system.
In this case, even though the system will eventually relax to a steady state, this steady state will be out-of-equilibrium due to persistent currents in the system.
In both cases, the degree to which the system is out of equilibrium at any given time is characterized by a positive rate of entropy production $\sigma_t > 0$.

Generically, a system may be driven by time-dependent and nonconservative forces at the same time.
Then, a natural question is whether the effects of both types of driving on the entropy production can be separated \cite{Lan78,Oon98,Hat01,Rue03,Kom08,Ber13,Mae14}.
For Brownian particles, we may decompose the entropy production rate into two nonnegative contributions, $\sigma_t = \sigma_t^\text{ex} + \sigma_t^\text{hk}$ called excess and housekeeping entropy production rate, respectively \cite{Hat01,Mae14}.
Here, $\sigma_t^\text{ex}$ is positive whenever the state of the system depends on time and vanishes in a steady state.
By contrast, $\sigma_t^\text{hk}$ is positive whenever the system is driven by a nonconservative force and vanishes if only conservative forces are acting on the system.
Somewhat surprisingly, this decomposition is not unique; specifically, the decompositions due to Hatano and Sasa \cite{Hat01} and due to Maes and Neto{\v{c}}n{\`y} \cite{Mae14} both satisfy the above properties but are generally distinct.

Such a decomposition of entropy production is very appealing from a theoretical point of view, since it allows deriving extended forms of fundamental results like the fluctuation theorem \cite{Hat01} and the Clausius heat theorem \cite{Mae14}.
However, the excess and housekeeping entropy are generally difficult to obtain directly from experimental and numerical data.
The reason is that, in the case of Ref.~\cite{Hat01}, we need to determine the instantaneous steady state of the system, whereas for Ref.~\cite{Mae14}, we need to construct a conservative force with the same time evolution, both of which typically require an analytic description of the system.
This issue has been addressed in Ref.~\cite{Kom08} by defining the excess entropy in terms of heat currents, with the downside that the result only holds in linear response.

In this Letter, our first main result is a geometric formalism for decomposing the entropy production rate into orthogonal gradient and nongradient fields, which describes both the Hatano-Sasa (HS) and the Maes-Neto{\v{c}}n{\`y} (MN) decomposition. 
This unifying formalism also provides the relation between the two decompositions: the MN decomposition is the one that minimizes the housekeeping part, which is thus always less than in the HS decomposition.
Our second main result is that, in the case of the MN decomposition, the geometric formalism provides variational expressions that can be used to calculate or estimate the excess and housekeeping entropy from experimental or numerical trajectory data.
This implies that the latter decomposition can be used to identify the contributions due to time-dependent and nonconservative driving in practical applications, while retaining its favorable theoretical properties.
We demonstrate our results using a particle in a time-dependent tilted periodic potential.

\begin{figure}
    \centering
    \includegraphics[width=\hsize]{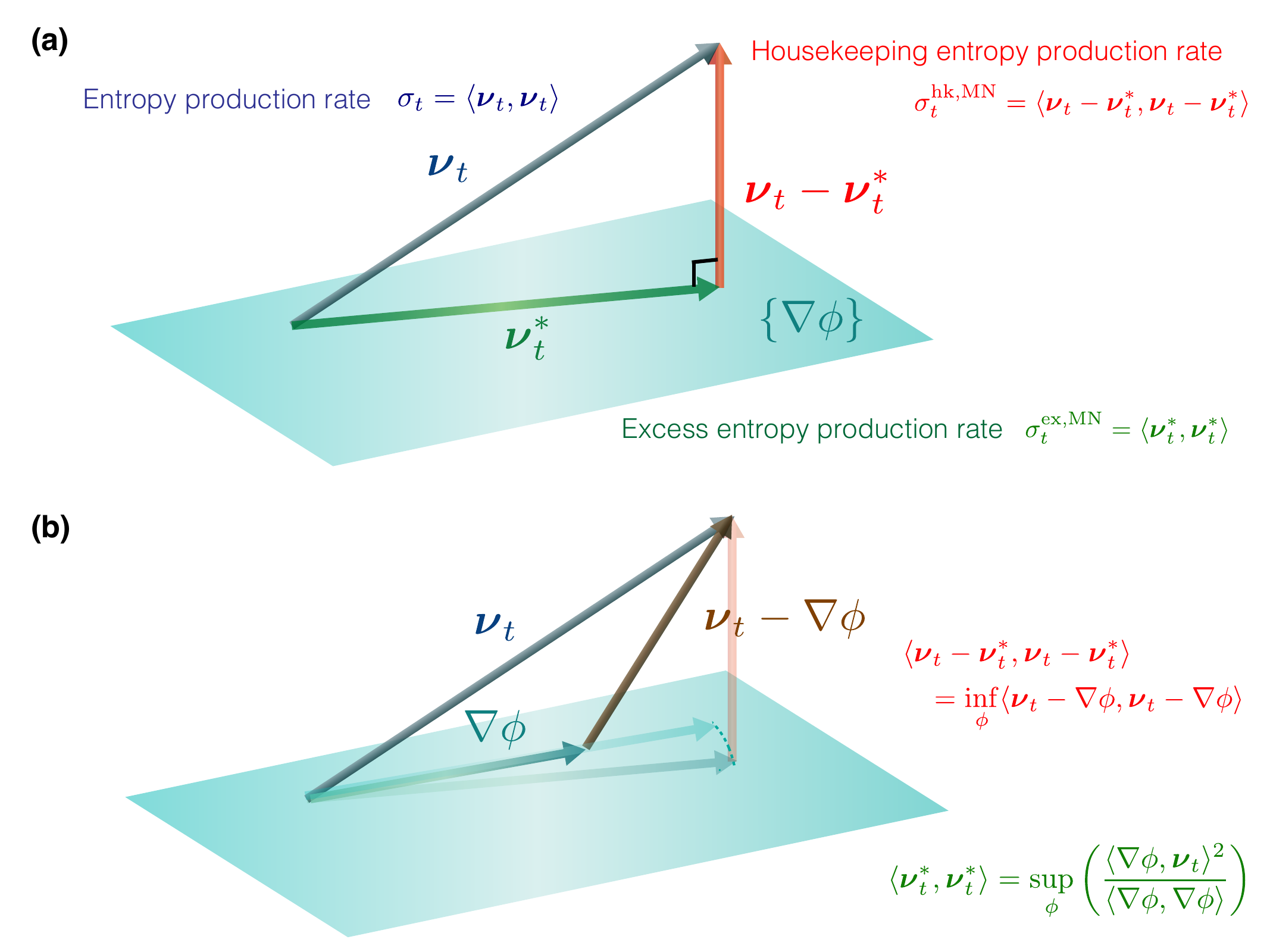}
    \caption{Geometric interpretations of the MN excess and housekeeping entropy production and their variational expressions. 
    (a) The velocity field $\bm{\nu}_t$ is decomposed into a gradient field $\bm{\nu}_t^*$ and its orthogonal complement $\bm{\nu}_t - \bm{\nu}_t^*$; the squared length of the two components yields the excess and housekeeping entropy production rate, respectively.
    (b) Since $\bm{\nu}_t^*$ is the orthogonal projection of $\bm{\nu}_t$ into the space of gradient fields $\lbrace \grad \phi \rbrace$, it can be characterized either by the gradient field $\grad \phi$ that maximizes the overlap with $\bm{\nu}_t$, leading to \eqref{variational-excess}, or by minimizing the length of the complement $\bm{\nu}_t - \grad \phi$, leading to \eqref{variational-housekeeping}. }
    \label{fig_geometry}
\end{figure}

\textit{Geometric decomposition of entropy production.}
The probability density $p_t$ of a system of Brownian particles with coordinates $\bm{x}(t)$ evolves according to the Fokker-Planck equation
\begin{subequations}
\begin{align}
\partial_t p_t(\bm{x}) &= - \grad \cdot \big(\bm{\nu}_t(\bm{x}) p_t(\bm{x}) \big) \qquad \text{with} \label{continuity}  \\
\bm{\nu}_t(\bm{x}) &= \mu \big( \bm{F}_t(\bm{x}) - T \grad \ln p_t(\bm{x}) \big) \label{meanvel} .
\end{align} \label{fokkerplanck}%
\end{subequations}
Here the time-dependent force $\bm{F}_t$ contains interactions between the particles as well as conservative and nonconservative external forces, $\mu$ is the particle mobility and $T$ is the temperature of the environment.
In the following, we will assume natural boundary conditions, that is, that the probability density and its derivatives vanish as $\Vert \bm{x} \Vert \rightarrow \infty$, where $\Vert \bm{x} \Vert$ denotes the Euclidean norm of $\bm{x}$.
The local mean velocity $\bm{\nu}_t$ is a vector field that describes the local average flows in the system.
Importantly, it also determines the rate of entropy production,
\begin{align}
\sigma_t = \av{\bm{\nu}_t,\bm{\nu}_t},
\end{align}
where we defined the inner product between two vector fields (assuming $\bm{u}$ and $\bm{v}$ are such that it exists)
\begin{align}
\av{\bm{u},\bm{v}} = \frac{1}{\mu  T} \int d\bm{x} \ \bm{u}(\bm{x}) \cdot \bm{v}(\bm{x}) p_t(\bm{x}) \label{inner-product} .
\end{align}
We then decompose the flows into two orthogonal components, 
\begin{align}
\bm{\nu}_t(\bm{x}) = \bm{\nu}^{(1)}_t(\bm{x}) + \bm{\nu}^{(2)}_t(\bm{x}) \quad \text{with} \quad \av{\bm{\nu}_t^{(1)},\bm{\nu}^{(2)}_t} = 0 \label{decomposition-orthogonal} .
\end{align}
For any such decomposition of the flows, the Pythagorean theorem immediately yields a decomposition of the entropy production rate into two positive parts,
\begin{align}
\sigma_t = \av{\bm{\nu}^{(1)}_t,\bm{\nu}^{(1)}_t} + \av{\bm{\nu}^{(2)}_t,\bm{\nu}^{(2)}_t} .
\end{align}
The decomposition \eqref{decomposition-orthogonal} is not unique; the goal is to find a physically meaningful decomposition.
We observe that, from the definition of the local mean velocity \eqref{fokkerplanck}, it can be written as a gradient field $\bm{\nu}_t = - \grad \psi_t$ whenever the forces acting in the system are conservative $\bm{F}_t = -\grad U_t$, where $U_t$ is the potential.
Since in this case, the housekeeping entropy production should vanish, we make the ansatz
\begin{align}
\bm{\nu}_t(\bm{x}) = - \grad \psi_t(\bm{x}) + \bar{\bm{\nu}}_t(\bm{x}) \label{decomposition-gradient},
\end{align}
that is, we decompose the local mean velocity into a gradient field and the remainder.
In the HS decomposition \cite{Hat01}, the central idea is to consider the instantaneous steady state $p_t^\text{st}$ of the system, which is attained when fixing the force $\bm{F}_t$ to its instantaneous value and letting the system relax to the corresponding steady state (which we assume to exist).
This steady state is characterized by a steady-state local mean velocity $\bm{\nu}_t^\text{st} = \mu (\bm{F}_t - T \grad \ln p_t^\text{st})$ which satisfies the steady state equation $\grad \cdot (\bm{\nu}_t^\text{st} p_t^\text{st}) = 0$.
Further, we have $\bm{\nu}_t - \bm{\nu}_t^\text{st} = - T \grad \ln (p_t/p_t^\text{st})$, which suggests choosing $\psi_t = T \ln(p_t/p_t^\text{st})$.
By explicit computation (see the Supplemental Material \cite{supmat}), it can be verified that this choice indeed satisfies \eqref{decomposition-orthogonal}.
We thus obtain the HS decomposition
\begin{align}
\sigma_t = \av{\bm{\nu}_t - \bm{\nu}_t^\text{st},\bm{\nu}_t - \bm{\nu}_t^\text{st}} + \av{\bm{\nu}_t^\text{st},\bm{\nu}_t^\text{st}} = \sigma_t^\text{ex,HS} + \sigma_t^\text{hk,HS} \label{HS-decomposition} .
\end{align}
Another possibility is to demand that $\bar{\bm{\nu}}_t$ should be orthogonal to all gradient fields, i.~e., $\av{\bar{\bm{\nu}}_t,\grad \phi} = 0$ for all $\phi$.
As we discuss below, this condition results in the MN decomposition \cite{Mae14}
\begin{align}
\sigma_t = \av{\bm{\nu}_t^*,\bm{\nu}_t^*} + \av{\bm{\nu}_t-\bm{\nu}_t^*,\bm{\nu}_t-\bm{\nu}_t^*} = \sigma_t^\text{ex,MN} + \sigma_t^\text{hk,MN} \label{MN-decomposition} ,
\end{align}
where $\bm{\nu}_t^*$ is the unique gradient field that satisfies $\grad \cdot (\bm{\nu}_t^* p_t) = \grad \cdot (\bm{\nu}_t p_t)$.
Thus, both the HS and MN decomposition can be viewed as decompositions of the local mean velocity $\bm{\nu}_t$ into a gradient field and an orthogonal remainder, which is our first main result.
This the geometrical intuition underlying the MN decomposition is illustrated in Fig.~\ref{fig_geometry}(a).

\textit{Variational expressions.}
Since the vector fields $\bm{\nu}_t^{(1)}$ and $\bm{\nu}_t^{(2)}$ in \eqref{decomposition-orthogonal} are orthogonal, they define a decomposition of the space $V$ of all local mean velocities into two orthogonal subspaces $V^{(1)}$ and $V^{(2)}$.
Conversely, $\bm{\nu}_t^{(1)}$ can be viewed as the orthogonal projection of $\bm{\nu}_t$ into the subspace $V^{(1)}$.
Then, we have two variational expressions for the square of the \enquote{length} of $\bm{\nu}_t^{(1)}$,
\begin{subequations}
\begin{align}
\av{\bm{\nu}_t^{(1)},\bm{\nu}_t^{(1)}} &= \sup_{\bm{v} \in V^{(1)}} \Bigg( \frac{\av{\bm{v},\bm{\nu}_t}^2}{\av{\bm{v},\bm{v}}} \Bigg) \\
& = \inf_{\bm{u} \in V^{(2)}} \big( \av{\bm{\nu}_t - \bm{u},\bm{\nu}_t - \bm{u}} \big) .
\end{align} \label{variational}%
\end{subequations}
The first expression follows by noting $\av{\bm{v},\bm{\nu}_t} = \av{\bm{v},\bm{\nu}^{(1)}_t}$ for all $\bm{v} \in V^{(1)}$ and then considering the equality condition of the Cauchy-Schwarz inequality $\av{\bm{v},\bm{\nu}^{(1)}_t}^2 \leq \av{\bm{v},\bm{v}} \av{\bm{\nu}_t^{(1)},\bm{\nu}_t^{(1)}}$.
The second expression can be confirmed by writing, for $\bm{u} \in V^{(2)}$,
\begin{align}
\av{\bm{\nu}_t - \bm{u},\bm{\nu}_t - \bm{u}} &= \av{\bm{\nu}_t^{(1)} + \bm{\nu}_t^{(2)} - \bm{u},\bm{\nu}_t^{(1)} + \bm{\nu}_t^{(2)} - \bm{u}} \nn
&= \av{\bm{\nu}_t^{(1)},\bm{\nu}_t^{(1)}} + \av{\bm{\nu}_t^{(2)} - \bm{u}, \bm{\nu}_t^{(2)} - \bm{u}},
\end{align}
which is minimized for $\bm{u} = \bm{\nu}_t^{(2)}$.
\eqref{variational} allows us to consider the inner product $\av{\bm{\nu}_t^{(1)},\bm{\nu}_t^{(1)}}$ either as a maximization over the subspace $V^{(1)}$ or a minimization over the subspace $V^{(2)}$.
For the MN decomposition, this is illustrated graphically in Fig.~\ref{fig_geometry}(b): The orthogonal projection can be obtained by either maximizing the overlap between $\bm{\nu}_t$ and some gradient field, or minimizing the length of the complement.

If $V^{(1)}$ is chosen as the space of gradient fields, then we immediately have
\begin{align}
\sigma_t^\text{ex,MN} = \inf_{\bm{u} \perp \grad \phi} \big(\av{\bm{\nu}_t - \bm{u}, \bm{\nu}_t - \bm{u}} \big) \label{MN-excess-variational} .
\end{align}
On the other hand, the orthogonality condition $\bm{u} \perp \grad \phi \Leftrightarrow \av{\bm{u},\grad \phi} = 0$ explicitly reads from \eqref{inner-product}
\begin{align}
0 &= \int d\bm{x} \ \grad \phi(\bm{x}) \bm{u}(\bm{x}) p_t(\bm{x}) \nn
& = - \int d\bm{x} \ \phi(\bm{x}) \grad \cdot \big(\bm{u}(\bm{x}) p_t(\bm{x}) \big), 
\end{align}
after integrating by parts.
Since the condition should hold for all $\phi$, this implies $\grad \cdot \big( \bm{u} p_t \big) = 0$.
Comparing this to \eqref{fokkerplanck}, the space $V^{(2)}$ can thus be characterized as all vector fields $\bm{u}$ that can be added to $\bm{\nu}_t$ without altering the time evolution of $p_t$.
Since this is equivalent to changing the force $\bm{F}_t$, \eqref{MN-excess-variational} implies a minimization of the entropy production rate with respect to the force, while keeping the time evolution of $p_t$ fixed.
This is exactly the minimum entropy production principle of Ref.~\cite{Mae14} and, thus, \eqref{MN-decomposition} is indeed the same as the MN decomposition.
In view of \eqref{variational}, the appealing feature of the MN decomposition is that one of the two subspaces has a simple mathematical characterization as the space of gradient fields, which allows us to write
\begin{align}
\sigma_t^\text{ex,MN} &= \sup_{\phi} \Bigg( \frac{\av{\grad \phi,\bm{\nu}_t}^2}{\av{\grad \phi, \grad \phi}} \Bigg), \label{variational-excess} \\
\sigma_t^\text{hk,MN} &= \inf_{\phi} \big( \av{\bm{\nu}_t - \grad \phi,\bm{\nu}_t - \grad \phi}  \big) \label{variational-housekeeping}.
\end{align}
From \eqref{variational-housekeeping}, we see that we may obtain an upper bound on the MN housekeeping entropy production rate by choosing an arbitrary gradient gradient field.
For the particular choice $\phi = -T \ln (p_t/p_t^\text{st})$, the right-hand side is equal to the HS housekeeping entropy production rate, so that we obtain the relation
\begin{align}
    \sigma_t^\text{hk,MN} \leq \sigma_t^\text{hk,HS} .
\end{align}
Thus, while the HS and MN decomposition are generally distinct, there exists a definite relation between the two.
We remark that, in principle, expressions similar to \eqref{variational-excess} and \eqref{variational-housekeeping} can be obtained for the HS decomposition; however, the structure of the orthogonal spaces is more complicated, and the resulting variational expressions are not convenient for practical applications.

\textit{Excess entropy and Wasserstein distance.}
In the following, we will focus on the MN decomposition and drop the superscript MN from now on.
Since the MN excess entropy production is the minimal entropy for a given time evolution of the probability density, we can also write is as
\begin{align}
    \sigma^\text{ex} = \inf_{\bm{\nu}_t \vert \partial_t p_t = -\grad \cdot (\bm{\nu}_t p_t)} \av{\bm{\nu}_t,\bm{\nu}_t},
\end{align}
where, on the right hand side, we minimize over the vector field $\bm{\nu}_t$ under the constraint that it satisfies the continuity equation \eqref{continuity}.
This expression closely resembles the Benamou-Brenier \cite{Ben00} formula from optimal transport theory.
Since the latter gives an equivalent expression of the Wasserstein distance between two probability densities \cite{Vil08}, we obtain the identification between the MN excess entropy production rate and the Wasserstein distance $\mathcal{W}$ (for more details, see \cite{supmat}),
\begin{align}
\sigma_t^\text{ex} = \frac{1}{\mu T} \lim_{\Delta t \rightarrow 0} \frac{\mathcal{W}(p_{t+\Delta t},p_t)^2}{\Delta t^2} \label{excess-Wasserstein}.
\end{align}
In Refs.~\cite{Aur11,Aur12,Dec19b}, it was found that the minimum entropy production associated with changing the probability density from an initial state $p_\text{i}$ to a finial state $p_\text{f}$ can be expressed in terms of the Wasserstein distance between the two states.
\eqref{excess-Wasserstein} generalizes this result to the case where, instead of the initial and final state, the time evolution of the probability density is fixed \cite{Nak21}.

\textit{Excess and housekeeping entropy from trajectory data.}
In order to obtain expressions more suited to applications, we use the explicit form of the inner product \eqref{inner-product},
\begin{align}
\av{\grad \phi,\bm{\nu}_t} &= \frac{1}{\mu T} \int d\bm{x} \ \grad \phi(\bm{x}) \cdot \bm{\nu}_t(\bm{x}) p_t(\bm{x}) \nn
&= \frac{1}{\mu T} \int d\bm{x} \ \phi(\bm{x}) \partial_t p_t(\bm{x}) = \frac{1}{\mu T} d_t \av{\phi}_t ,
\end{align}
where we integrated by parts and used \eqref{fokkerplanck}.
Here $\av{\phi}_t$ denotes an average with respect to $p_t$.
This allows us to write the excess entropy production rate as
\begin{align}
\sigma_t^\text{ex} &= \frac{1}{\mu T} \sup_{\phi} \Bigg( \frac{\big(d_t \av{\phi}_t\big)^2}{\av{\Vert \grad \phi \Vert^2}_t} \Bigg) \label{variational-excess-2} .
\end{align}
The right-hand side can be evaluated by only considering scalar observables $\phi$ that are a function of the position $\bm{x}(t)$ of the Brownian particles.
The maximization can then be readily performed using a suitable parameterization of $\phi$.
The maximizer $\phi^*$ of \eqref{variational-excess-2} yields the optimal local mean velocity up to a constant, $\bm{\nu}_t^* = -c \grad \phi^*$.
This is the flow field that yields the same time evolution as \eqref{fokkerplanck} while minimizing the entropy production rate.
In contrast to directly minimizing the entropy production rate, \eqref{variational-excess-2} does not require any additional constraints.
Instead of maximizing the right-hand side of \eqref{variational-excess-2} with respect to $\phi$, we can also choose an arbitrary scalar function and obtain a lower bound.
We remark that \eqref{variational-excess} is closely related to the short-time version of the thermodynamic uncertainty relation \cite{Man20,Ots20,Vu20}: If the maximization is taken over all vector fields, then the result is the total entropy production rate; by restricting the maximization to gradient fields, the result is the excess entropy production.

In order to obtain a similar expression for the housekeeping entropy, we write the force acting on the system as $\bm{F}_t = -\grad U_t + \bm{F}_t^\text{nc}$, where $\bm{F}_t^\text{nc}$ is a nonconservative force.
Since in \eqref{variational-housekeeping}, the infimum is taken over all gradient fields, we may absorb the gradient terms in the local mean velocity into $V = \phi/\mu + T \ln p_t + U_t$ and write
\begin{align}
\sigma_t^\text{hk} = \frac{\mu}{T} \inf_{V} \Av{\big\Vert \bm{F}_t^\text{nc} - \grad V \big\Vert^2}_t. \label{variational-housekeeping-2}
\end{align}
In many physical settings, the nonconservative force is an externally applied driving force and its functional form is therefore known.
In such cases, we can evaluate \eqref{variational-housekeeping-2} by considering only scalar observables $V$ that depend on the position of the particles.
We stress that \eqref{variational-housekeeping-2} does not depend explicitly on the potential $U_t$, which generally includes interactions between particles and is therefore often not known precisely in practice.
Without minimizing, an arbitrary choice of $V$ yields an upper bound on the housekeeping entropy production rate.
One meaningful such choice is $V = \av{\bm{F}_t^\text{nc}}_t \cdot \bm{x}$, which yields the upper bound
\begin{align}
\sigma_t^\text{hk} \leq \frac{\mu}{T} \Av{\big\Vert \bm{F}_t^\text{nc} \big\Vert^2} - \Big\Vert \Av{\bm{F}_t^\text{nc}}_t \Big\Vert^2 = \frac{\mu}{T} \text{Var}(\bm{F}_t^\text{nc}).
\end{align}
Thus, the housekeeping entropy production rate is bounded by the variance of the nonconservative force.
In summary, the variational expressions \eqref{variational-excess-2} and \eqref{variational-housekeeping-2} allow us to determine both the excess and the housekeeping entropy production rate from given trajectory data, which is our second main result.

\begin{figure}
    \centering
    \includegraphics[width=\hsize]{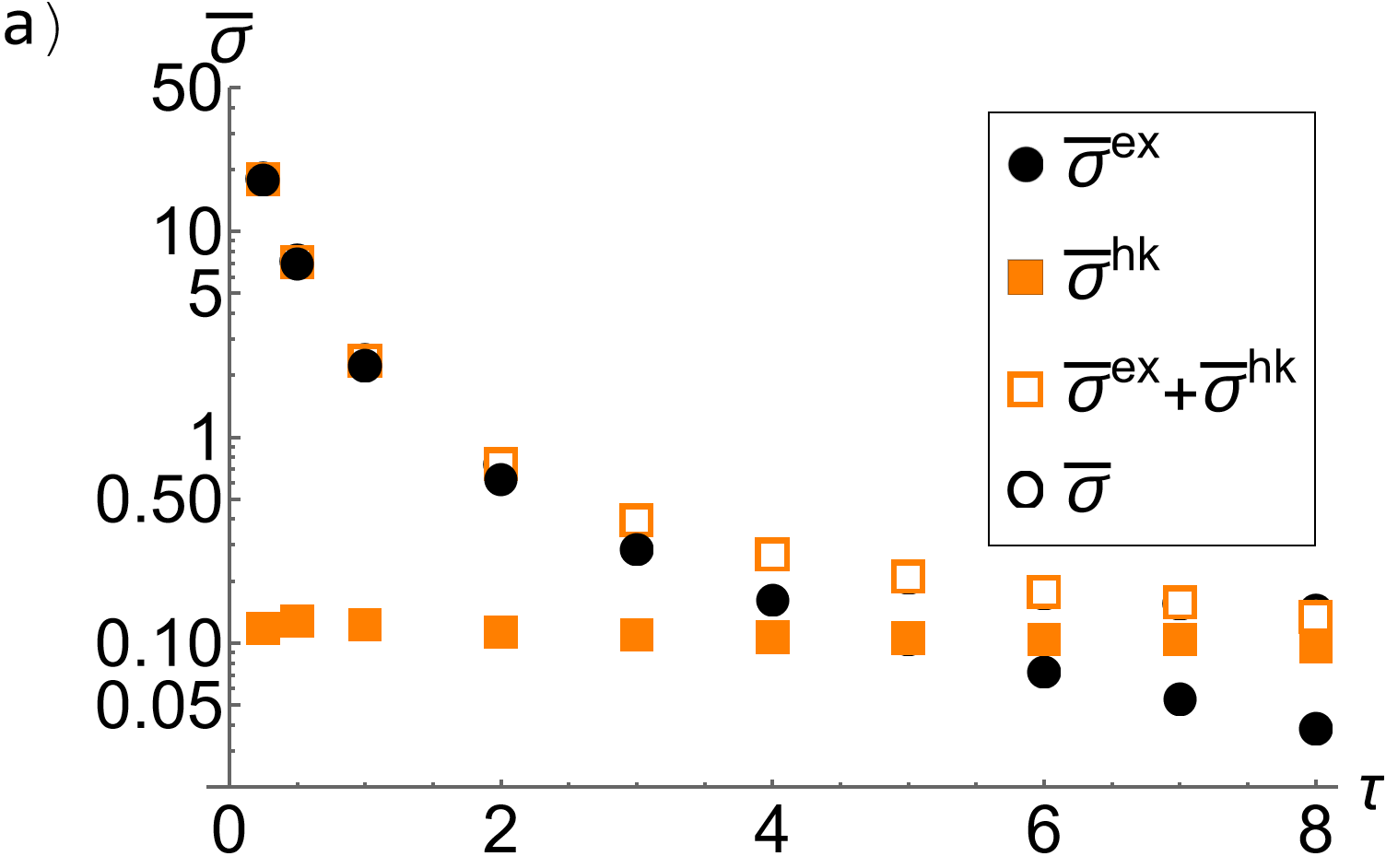}
    \includegraphics[width=\hsize]{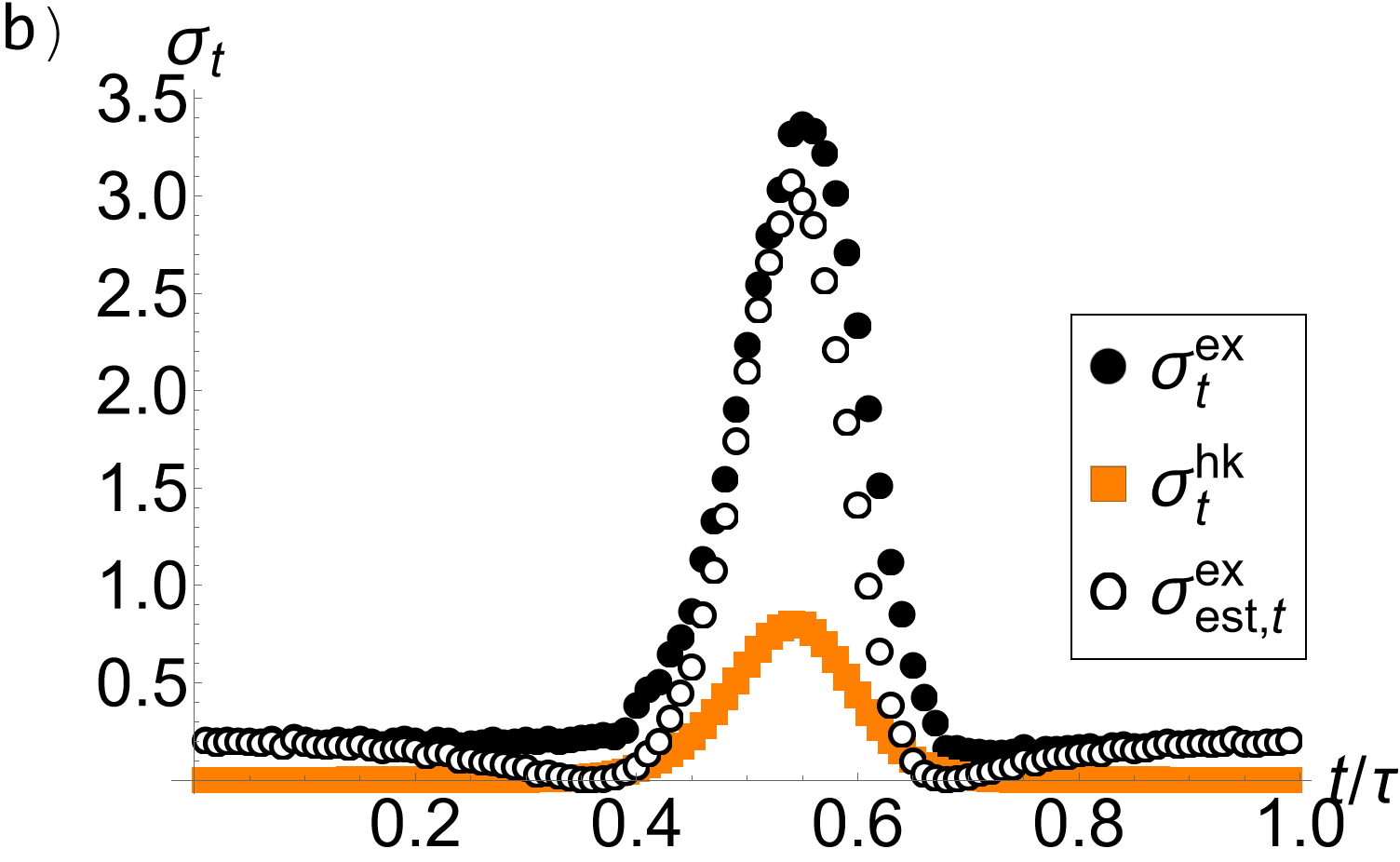}
    \caption{The excess and housekeeping entropy production rate for the dynamics \eqref{langevin-perpot} with the potential \eqref{potential}. Parameters used for the simulation are $U_0 = 1$, $T = 0.25$, $\mu = 1$, $L = 1$  and $F^\text{nc}_0 = 0.25$. We used a total of $50$ periods for $10000$ trajectories and $K = 10$ modes for the minimization in \eqref{phi-parameterization}. a) The time-averaged entropy production rates $\bar{\sigma} = \int_0^\tau dt \ \sigma_t/\tau$ as a function of the driving period $\tau$.  b) The instantaneous entropy production rates as a function of time for $\tau = 2$. We also show the lower bound $\sigma^\text{ex}_{\text{est},t}$ obtained by choosing $\phi(x) = \cos(\lambda x)$ in \eqref{variational-excess-2}. }
    \label{fig_entropy}
\end{figure}

\textit{Demonstration.}
As an explicit demonstration of our previous results, we study the motion of a Brownian particle in a time-dependent periodic potential $U_t(x+L) = U_t(x)$, which is driven by a constant bias $F^\text{nc}_0$,
\begin{align}
    \dot{x}(t) = \mu \big(- \partial_x U_t(x(t)) + F^\text{nc}_0 \big) + \sqrt{2 \mu  T} \xi(t) \label{langevin-perpot} .
\end{align}
Such a situation is common in experimental systems to study the dynamics of colloidal particles \cite{Spe07,Evs08}.
For simplicity, we modulate the potential in a time-periodic manner, $U_{t+\tau}(x) = U_t(x)$.
For long times, the probability density is then periodic in both space and time, $p_{t+\tau}(x) = p_{t}(x) = p_t(x+L)$.
We perform numerical simulations of \eqref{langevin-perpot}, from which we obtain a set of trajectories, which we then use to compute the excess and housekeeping entropy production rate according to \eqref{variational-excess-2} and \eqref{variational-housekeeping-2}.
Note that for periodic boundary conditions, only forces that can be written as the gradient of a \textit{periodic} scalar function are conservative.
Thus, we parameterize the scalar functions $V(x)$ and $\phi(x)$ as 
\begin{align}
   V(x) = \sum_{k = 0}^K \big( a_k \sin(k \lambda x) + b_k \cos(k \lambda x) \big) \label{phi-parameterization},
\end{align}
with $\lambda = 2\pi/L$ and evaluate \eqref{variational-excess-2} and \eqref{variational-housekeeping-2}.
Then, we numerically optimize resulting expressions with respect to the parameters $a_k$ and $b_k$ using Mathematica's \texttt{NMaximize} and \texttt{NMinimize} routines.
We stress that determining $\sigma_t^\text{ex}$ and $\sigma_t^\text{hk}$ in this manner requires only the trajectories and the value of the bias $F_0$.
As a concrete example, we choose the space-time periodic potential
\begin{align}
    U_t(x) = U_0 \big( \sin(\lambda x) + A \sin(\lambda x - \omega t) \big) \label{potential} ,
\end{align}
which corresponds to a sine-shaped potential with a time-dependent component of amplitude $A$ and period $\tau = 2\pi/\omega$.
The resulting excess and housekeeping entropy rates averaged over one period are shown as a function of the driving period $\tau$ in Fig.~\ref{fig_entropy}a).
First, we note that the sum $\sigma_t^\text{ex} + \sigma_t^\text{hk}$ precisely reproduces the entropy production rate, calculated according to the stochastic thermodynamics result $\sigma_t = \av{F_t \circ \dot{x}}/T$ \cite{Sek10,Sei12}.
As expected, the entropy is dominated by the excess contribution for fast driving, while for slow driving, the housekeeping part from the constant bias is dominant.
For the present example, the housekeeping entropy rate is almost independent of the driving speed, reflecting that, to a good approximation, the time-dependent probability density depends on $\tau$ only via a rescaling of time.
However, unlike in a steady state, the housekeeping entropy production rate strongly depends on time, as can be seen from Fig.~\ref{fig_entropy}b): The main contribution to the entropy production stems from times $t \approx \tau/2$, where the total depth of the potential is minimal.

\textit{Discussion.}
In this work, we decomposed the entropy production using the geometric formalism of orthogonal projections.
While the extension to diffusion matrices and multiplicative noise is possible, a more serious challenge is to find a similar interpretation for other classes of stochastic dynamics, notably underdamped Langevin and Markov jump dynamics \cite{Mae07}.
In both cases, it has been recently shown that entropy production is bounded from below by an appropriately defined Wasserstein distance \cite{Dec19b,Van21,Dec21c}; in light of \eqref{excess-Wasserstein}, we may thus speculate that the latter can be identified with an excess entropy similar to the MN decomposition also in these cases.

Generally, variational principles and geometry are often intimately connected, be it in classical mechanics \cite{Arn13} or optimal transport theory \cite{Vil08}.
The geometric decomposition \eqref{decomposition-gradient} implies the variational formulas \eqref{variational} for the individual contributions to the entropy production rate.
Recently, the partial entropy production of subsystems was also shown to follow from a variational principle \cite{ito2020unified}, based on the projection theorem of information geometry \cite{amari2016information}.
While the maximum entropy principle of equilibrium statistical mechanics \cite{Jay82} can be recast in a geometric formalism involving an orthogonal decomposition of the underlying space \cite{Pav13}, the above results suggest that a viable for obtaining minimum entropy production principles \cite{Jay80} may be starting from a suitable geometric decomposition of the space.

\begin{acknowledgments}
S.~I.~thanks Muka Nakazato, Masafumi Oizumi, Shin-ichi Amari and Kohei Yoshimura for fruitful discussions. 
S.~I.~is supported by JSPS KAKENHI (Grant No. 19H05796, 21H01560), JST Presto (Grant No. JPMJPR18M2) and UTEC-UTokyo FSI Research Grant Program.
S.~S.~is supported by JSPS KAKENHI (Grant No. 17H01148, 19H05795, and 20K20425).
\end{acknowledgments}

\def\theequation{S\arabic{equation}}

\section*{Supplemental Material}

\subsection{Orthogonality for the Hatano-Sasa decomposition}
In the Hatano-Sasa decomposition of the entropy production rate, the local mean velocity is decomposed using the local mean velocity of the instantaneous steady state
\begin{align}
    \bm{\nu}_t(\bm{x}) = \bm{\nu}_t^\text{st}(\bm{x}) + \bm{\nu}_t(\bm{x}) - \bm{\nu}_t^\text{st}(\bm{x}) \label{HS-meanvel},
\end{align}
where $\bm{\nu}_t^\text{st}$ satisfies
\begin{align}
    \bm{\nu}_t^\text{st}(\bm{x}) &= \mu \big(\bm{F}_t(\bm{x}) - T \grad \ln p_t^\text{st}(\bm{x}) \big) \quad \text{with} \label{steady-state} \\
    0 &= - \grad \cdot \big( \bm{\nu}_t^\text{st}(\bm{x}) p_t^\text{st}(\bm{x}) \big) \n .
\end{align}
Comparing this to the definition of the local mean velocity (see Eq.~(1) of the main text),
\begin{align}
\bm{\nu}_t(\bm{x}) = \mu \big(\bm{F}_t(\bm{x}) - T \grad \ln p_t(\bm{x}) \big),
\end{align}
we see that the difference $\bm{\nu}_t - \bm{\nu}_t^\text{st}$ is a gradient field,
\begin{align}
    \bm{\nu}_t(\bm{x}) - \bm{\nu}_t^\text{st}(\bm{x}) = - \mu T \grad \ln \bigg( \frac{p_t(\bm{x})}{p_t^\text{st}(\bm{x})} \bigg),
\end{align}
so that this decomposition is of the form (see Eq.~(6) of the main text)
\begin{align}
\bm{\nu}_t(\bm{x}) = - \grad \psi_t(\bm{x}) + \bm{u}_t(\bm{x}) .
\end{align}
Next, we calculate the inner product (defined in Eq.~(3) of the main text) between $\bm{\nu}_t - \bm{\nu}_t^\text{st}$ and $\bm{\nu}_t^\text{st}$
\begin{align}
    \av{\bm{\nu}_t - \bm{\nu}_t^\text{st},\bm{\nu}_t^\text{st}} = - \int d\bm{x} \ \grad \ln \bigg( \frac{p_t(\bm{x})}{p_t^\text{st}(\bm{x})} \bigg) \cdot \bm{\nu}_t^\text{st}(\bm{x}) p_t(\bm{x}) .
\end{align}
We introduce a factor $1 = p_t^\text{st}/p_t^\text{st}$ and note that $f \grad \ln f = \grad f$ to obtain
\begin{align}
    \av{\bm{\nu}_t - \bm{\nu}_t^\text{st},\bm{\nu}_t^\text{st}} = - \int d\bm{x} \ \grad \bigg( \frac{p_t(\bm{x})}{p_t^\text{st}(\bm{x})} \bigg) \cdot \bm{\nu}_t^\text{st}(\bm{x}) p_t^\text{st}(\bm{x}) .
\end{align}
Using the divergence theorem and \eqref{steady-state}, we find
\begin{align}
    \av{\bm{\nu}_t - \bm{\nu}_t^\text{st},\bm{\nu}_t^\text{st}} = 0.
\end{align}
Thus, the two terms in \eqref{HS-meanvel} are indeed orthogonal with respect to the inner product defined by $p_t$.

\subsection{Excess entropy and Wasserstein distance}
As discussed in the main text and in Ref.~\cite{Mae14}, the Maes-Neto{\v{c}}n{\`y} excess entropy production is the minimum entropy production rate associated with the time evolution of the probability density $p_t$.
In Refs.~\cite{Aur11,Aur12,Dec19b} it was found that the minimum entropy production associated with changing the probability density from an initial state $p_\text{i}$ to a finial state $p_\text{f}$ can be expressed in terms of the Wasserstein distance $\mathcal{W}(p_\text{f},p_\text{i})$ \cite{Vil08} between the two states
\begin{align}
\Delta S^\text{min} = \frac{1}{\mu T \tau} \mathcal{W}(p_\text{f},p_\text{i})^2 \label{minent-Wasserstein} ,
\end{align}
where $\tau$ is the duration of the process.
For an arbitrary process connecting the two states, the right-hand side is a lower bound on the entropy production $\Delta S$.
For a given time evolution connecting $p_\text{i}$ to $p_\text{f}$, we can imagine minimizing the entropy production rate at any instant of time.
Since the result is still a process connecting the same initial and final state, we immediately have
\begin{align}
\Delta S^\text{ex} \geq \frac{1}{\mu T \tau} \mathcal{W}(p_\text{f},p_\text{i})^2 \label{excess-Wasserstein-bound},
\end{align}
where $\Delta S^\text{ex} = \int_0^\tau dt \ \sigma_t^\text{ex}$ is the excess entropy production.
This implies that the right-hand side of \eqref{minent-Wasserstein} can estimate only the excess part of the entropy production.
Further, in Ref.~\cite{Nak21} it was shown that 
\begin{align}
\sigma_t \geq \frac{1}{\mu T} \lim_{\Delta t \rightarrow 0} \frac{\mathcal{W}(p_{t+\Delta t},p_t)^2}{\Delta t^2} \label{entropy-Wasserstein-inequality},
\end{align}
with equality when the dynamics is driven by a conservative force.
Since the excess entropy production rate represents a process with the same time evolution and driven by a conservative force, we immediately have the identification
\begin{align}
\sigma_t^\text{ex,MN} = \frac{1}{\mu T} \lim_{\Delta t \rightarrow 0} \frac{\mathcal{W}(p_{t+\Delta t},p_t)^2}{\Delta t^2} ,
\end{align}
which shows that \eqref{excess-Wasserstein-bound} becomes an equality in the short-time limit.
Thus, we can identify the Maes-Neto{\v{c}}n{\`y} excess entropy production rate with the infinitesimal Wasserstein distance along the time evolution of the probability density.

\end{document}